\documentclass[aps,twocolumn,showpacs,preprintnumbers,amsmath,amssymb,superscriptaddress,floatfix,nofootinbib]{revtex4}
\usepackage[colorlinks=true]{hyperref}
\usepackage{xcolor}
\hypersetup{colorlinks=true, citecolor=blue, urlcolor=blue, linkcolor=blue}
\usepackage{graphicx}
\usepackage{epsfig}
\usepackage{epstopdf}
\usepackage{hyperref}
\usepackage{amsmath}
\usepackage{amsfonts}
\usepackage{amssymb}

\usepackage{natbib}

\usepackage{graphicx,color,dcolumn,booktabs}
\usepackage{txfonts}
\usepackage{amssymb}
\usepackage{indentfirst}
\usepackage{subfig}
\usepackage{float}
\usepackage{multirow}
\usepackage{color}

\allowdisplaybreaks
\begin{document}

\title{Explanation of the newly obseaved $Z_{cs}^-(3985)$ as a $D_s^{(*)-}D^{(*)0}$ molecular state}


\author{Zhi-Feng Sun}
\affiliation{School of Physical Science and Technology,
Lanzhou University, Lanzhou 730000, China}

\author{Chu-Wen Xiao}
\email{xiaochw@csu.edu.cn (Coresponding Author)}
\affiliation{School of Physics and Electronics,
Central South University, Changsha 410083, China}

\date{\today}

\begin{abstract}
Inspired by the newly observed $Z_{cs}^-(3985)$ by BESIII collaboration, we study the structure of this particle in the picture of $D_s^{(*)-}D^{(*)0}$ molecular state. Firstly we systematically construct the Lagrangians which describing the interaction of charmed mesons, taking into account the chiral and hidden local symmetries. With the obtained effective potentials from the Lagrangians constructed, we solve the coupled channel Bethe-Salpeter equation with the on-shell approximation. On the third Reimann sheet, a pole position of around $3982.34-i0.53$ MeV is obtained, which can be associated to the $Z_{cs}^-(3985)$ and explained as a loose bound state of $D_s^*\bar{D}^*$.
\end{abstract}

\maketitle

\section{Introduction}
After the observation of $X(3872)$ by Belle collaboration in 2003 \cite{Choi:2003ue}, a series of charmonium-like states have been discovered, which have caught much attention by both theorists and experimentalists. These particles are difficult to understood under the picture of charmonium, while the explanation of multiquark states becomes a general proposal.

In 2013, BESIII collaboration reported their observations of the charged $Z_c^+(3900)$ \cite{Ablikim:2013mio}. Soon after that, the Belle collaboration confirmed the existence of this particle \cite{Liu:2013dau}. Using the data collected by CLEO-c collaboration, T. Xiao {\it et al.} confirmed the existence of $Z_c^+(3900)$ and also discovered its neutral partner $Z_c^0(3900)$ \cite{Xiao:2013iha}. The follow-up experiment by BESIII also observed the neutral $Z_c^0(3900)$ \cite{Ablikim:2015tbp}. In 2017, BESIII determined the spin-parity of $Z_c(3900)$ as $1^+$ \cite{Collaboration:2017njt}. D0 collaboration measured the signal of $Z_c(3900)\to \pi^\pm J/\psi$ semi-inclusive weak decays of $b$ flavored hadrons \cite{Abazov:2018cyu}. The mass and width of $Z_c(3900)$ from Particle Data Group (PDG) is given by \cite{PDG}
\begin{eqnarray}
m=3887.2 \pm 2.3 \ {\rm MeV}, \ \Gamma=28.2 \pm 2.6 \ {\rm MeV}.
\end{eqnarray}
It is worth mention that, BESIII discovered a charged and a neutral states in $e^+e^-\to [D\bar{D}^*+c.c]^{\pm,0}\pi^{\mp,0}$, i.e., $Z_c(3885)$ \cite{Ablikim:2013xfr}. Since the resonance parameters of $Z_c(3885)$ and $Z_c(3900)$ are close to each other, these two particles are probably the same one.

Meanwhile, BESIII experiment also explored other charmonium-like states. In the process of $e^+e^-\to \pi^+\pi^-h_c$ with the energy of $3.90-4.42$ GeV, the signal of the charged $Z_c^+(4020)$ was discovered \cite{Ablikim:2013wzq}. Later, the neutral partner of $Z_c^+(4020)$ was observed by BESIII in the channel of $e^+e^-\to \pi^0\pi^0h_c$ \cite{Ablikim:2014dxl}.
The cross section of $e^+e^-\to \pi^0\pi^0h_c$ process is half of that of $e^+e^-\to \pi^+\pi^-h_c$, which prove that the $Z_c(4020)$ is an isovector particle \cite{Yuan:2015kya}. On the other hand, BESIII reported the neutral and charged $Z_c(4025)$ in $e^+e^-\to [D^*\bar{D}^*]^\pm \pi^0$ and $e^+e^-\to D^{*0}\bar{D}^{*0}(D^{*+}{D}^{*-})\pi^0$ channels \cite{Ablikim:2013emm}, whose resonance parameters are consistent with those of $Z_c(4020)$. So it is reasonable to identify that these two particles are the same state. The mass and width of $Z_c(4020)$ listed in PDG are \cite{PDG}
\begin{eqnarray}
m=4024.1\pm 1.9\ {\rm MeV},\ \Gamma=13\pm 5\ {\rm MeV}.
\end{eqnarray}

After the observation of these particles, many groups tried to study the nature of them. The theoretical explanations involve the hadronic molecules \cite{Chen:2015ata, Dong:2013iqa, He:2013nwa}, the compact tetraquark states \cite{Kisslinger:2014mwa, Agaev:2016dev, Agaev:2017tzv}, the kinematical effects \cite{Swanson:2014tra, Chen:2013coa, Szczepaniak:2015eza} and so on.

Except for the results from BESIII, the other experimental collorations also made efforts to search for the charmonium-like states. For instance, the $Z_c(4200)$ was observed by Belle \cite{Chilikin:2014bkk}, and $X(4500)$, $X(4700)$ by LHCb \cite{Aaij:2016iza}. The study of charmonium-like states will deepen our understandings of the structure of new hadronic states and the phenomenon in the low energy region of QCD.

Besides the charmonium-like states, searching for the $Z_{cs}$ states with the quark components of $c$, $\bar{c}$, $\bar{s}$ and $q$ (refer to the light ones of $u$ and $d$) also catch much theoretical attentions. In principle, the $Z_c$ states, assumed to be composed with $c$, $\bar{c}$, $\bar{q}$ and $q$, should have their SU(3) flavour partner of the strangeness, which inspires us to study the $Z_{cs}$ states. 

Very recently, BESIII collaboration reports new results in the $e^+e^-\to K^+ (D_s^-D^{*0}+D_s^{*-}D^0)$ process.  An excess over the known contributions of the conventional charmed mesons is observed near the $D_s^-D^{*0}$ and $D_s^{*-}D^0$ mass thresholds in the $K^+$ recoil-mass spectrum for the events collected at $\sqrt{s}= 4.681$ GeV \cite{Zcs}. The corresponding mass and width are determined as
\begin{eqnarray}
m=3982.5^{+1.8}_{-2.6}\pm 2.1\ {\rm MeV}, \ \Gamma=12.8^{+5.3}_{-4.4}\pm 3.0\ {\rm MeV}.
\end{eqnarray}
This exiting discovery enriches our understandings of the spectrum of hadronic states, and much attention will be paid in this field.

In Ref. \cite{1830601}, Wang et al. propose a novel method to detect higher charmed/charmed-strange meson. In Ref. \cite{1830623}, the charged $Z_{cs}(3985)$ structure under a reﬂection mechanism is proposed. In Ref. \cite{1830580}, the authors perform the $J/\pi (K)$, $D_{(s)}\bar{D}^*$, $D_{(s)}^*\bar{D}$ and $D_{(s)}^*\bar{D}^*$ coupled-channel calculation in the contact interaction with the SU(3) flavour symmetry and heavy quark spin symmetry, and the obtained mass and width of $Z_{cs}(3985)$ are consistent with the experimental data.
In Ref. \cite{1830608}, using one-boson-exchange model, the authors point out that the $Z_{cs}(3985)$ can not be a $\bar{D}_s^*D$ and $\bar{D}D^*$ bound state. In Ref. \cite{1830582}, Yang et al. point out that the $Z_{cs}$ state could be either a virtual state or a resonance. Using QCD sum rule \cite{1830632}, the authors find that the configuration of $Z_{cs}(3985)$ could be a mixture of different structures.

In this work, inspired by the new observation by BESIII collaboration, we will study whether or not the newly observed $Z_{cs}$ can be interpreted as a molecular state considering the coupled channels of $D_s\bar{D}^*$, $D_s^*\bar{D}$ and $D_s^*\bar{D}^*$. The paper is organized as follows. After this introduction, we will show constructed Lagrangians in Section II. In Section III and IV we calculate the effective potentials and solve the on-shell Bethe-Salpeter equation. We will give the numerical results in Section V. Finally, a short summary is given.

\section{Lagrangians}
According to the chiral symmetry \cite{Cheng:1992xi,Yan:1992gz} and the hidden local symmetry \cite{Bando:1984ej,Bando:1987br, Meissner:1987ge, Harada:2003jx}, the Lagrangians are constructed describing the interactions of two charmed mesons and a light meson. In the following, we list the Lagrangians up to order 1,
\begin{eqnarray}
\mathcal{L}_1&=&a_1(iP\hat{\alpha}_{\|\mu }D^\mu P^\dag+h.c.)+b_1(P\hat{\alpha}_{\bot \mu}P^{*\mu \dag}+h.c.)\nonumber\\
&&+c_2(\epsilon^{\mu\nu\alpha\beta}P^*_\nu\hat{\alpha}_{\bot\alpha}D_\mu P^{*\dag}_\beta+h.c.)+c_3(iP^*_\nu\hat{\alpha}^\mu_\|D_\mu P^{*\nu\dag}\nonumber\\
&&+h.c.),\label{eq1}
\end{eqnarray}
where
\begin{eqnarray}
V_\mu&=&\frac{g}{\sqrt{2}}\left(\begin{array}{ccc}
\frac{1}{\sqrt{2}}(\rho^0+\omega)&\rho^+ &K^{*+}\\
\rho^-&-\frac{1}{\sqrt{2}}(\rho^0-\omega)&K^{*0}\\
K^{*-}&\bar{K}^{*0}&\phi
\end{array}\right)_\mu,\\
\Phi&=&\left(\begin{array}{ccc}
\frac{\sqrt{3}\pi^0+\eta_8+\sqrt{2}\eta_0}{\sqrt{3}}&\sqrt{2}\pi^+ &\sqrt{2}K^{+}\\
\sqrt{2}\pi^-&\frac{-\sqrt{3}\pi^0+\eta_8+\sqrt{2}\eta_0}{\sqrt{3}}&\sqrt{2}K^{0}\\
\sqrt{2}K^{-}&\sqrt{2}\bar{K}^{0}&\frac{-2\eta_8+\sqrt{2}\eta_0}{\sqrt{3}}
\end{array}\right),
\\
\hat{\alpha}_{\bot\mu}&=&(D_\mu \xi_R\xi_R^\dag-D_\mu \xi_L\xi_L^\dag)/(2i),\\
\hat{\alpha}_{\|\mu}&=&(D_\mu \xi_R\xi_R^\dag+D_\mu \xi_L\xi_L^\dag)/(2i),\\
\xi_L &=& e^{i\sigma/F_\sigma}e^{-i\Phi/(2F_\pi)},\\
\xi_R &=& e^{i\sigma/F_\sigma}e^{i\Phi/(2F_\pi)},\\
D_\mu\xi_L&=&\partial_\mu\xi_L-iV_\mu\xi_L+i\xi_Ll_\mu,\\
D_\mu\xi_R&=&\partial_\mu\xi_R-iV_\mu\xi_R+i\xi_Rr_\mu,\\
P&=&(D^0,D^+,D_s^+)\ \mathbf{or} \ (B^-,\bar{B}^0,\bar{B}_s^0),\\
P^*_\tau &=&(D^{*0},D^{*+},D_s^{*+})_\tau\ \mathbf{or} \ (B^{*-},\bar{B}^{*0},\bar{B}_s^{*0})_\tau,\\
D_\mu P&=&\partial_\mu P+iP\alpha^\dag _{\| \mu}=\partial_\mu P+iP\alpha_{\| \mu},\\
D_\mu P^*_\tau&=&\partial_\mu P^*_\tau+iP^*_\tau\alpha^\dag _{\| \mu}=\partial_\mu P^*_\tau+iP^*_\tau\alpha_{\| \mu},
\end{eqnarray}
with $F_\pi=93$ MeV. Note that in the unitary gauge, $\sigma=0$, i.e., $\xi_L=\xi_R=e^{-i\Phi/(2F_\pi)}$.

The contact terms involving four charmed mesons are also constructed, i.e.,
\begin{eqnarray}
\mathcal{L}&=&d_1(PP^\dag)(PP^\dag)+d_2(PP^\dag)(P^{*\mu}P^{*\dag}_\mu)\nonumber\\
&&+d_3\{(PP^{*\dag}_\mu)(PP^{*\mu\dag})+(P^{*}_\mu P^\dag)(P^{*\mu}P^\dag)\}\nonumber\\
&&+d_4(PP^{*\dag}_\mu)(P^{*\mu}P^\dag)\nonumber\\
&&+d_5\{i\epsilon_{\mu\nu\alpha\beta}(D^\mu PP^{*\nu\dag})(P^{*\alpha}P^{*\beta\dag})\nonumber\\
&&-i\epsilon_{\mu\nu\alpha\beta}(P^{*\nu}D^\mu P^\dag)(P^{*\beta}P^{*\alpha\dag})\}\nonumber\\
&&+d_6\{i\epsilon_{\mu\nu\alpha\beta}( PD^\mu P^{*\nu\dag})(P^{*\alpha}P^{*\beta\dag})\nonumber\\
&&-i\epsilon_{\mu\nu\alpha\beta}(D^\mu P^{*\nu}P^\dag)(P^{*\beta}P^{*\alpha\dag})\}\nonumber\\
&&+d_7\{i\epsilon_{\mu\nu\alpha\beta}(PP^{*\nu\dag})(D^\mu P^{*\alpha}P^{*\beta\dag})\nonumber\\
&&-i\epsilon_{\mu\nu\alpha\beta}(P^{*\nu}P^\dag)(P^{*\beta}D^\mu P^{*\alpha\dag})\}\nonumber\\
&&+d_8 \{i\epsilon_{\mu\nu\alpha\beta}(PP^{*\nu\dag})(P^{*\alpha}D^\mu P^{*\beta\dag})\nonumber\\
&&-i\epsilon_{\mu\nu\alpha\beta}(P^{*\nu}P^\dag)(D^\mu P^{*\beta}P^{*\alpha\dag})\}\nonumber\\
&&+d_9(P^{*\mu}P_\mu^{*\dag})(P^{*\nu}P_\nu^{*\dag})+d_{10}(P^{*\mu}P_\nu^{*\dag})(P^{*}_\mu P^{*\nu\dag})\nonumber\\
&&+d_{11}(P^{*\mu}P_\nu^{*\dag})(P^{*\nu}P_\mu^{*\dag}),
\end{eqnarray}
with the coefficients of $d_i$.

\section{Effective potentials with partial wave projection}
With the constructed Lagrangians above, we obtain the amplitudes of the contact terms with isospin $\frac{1}{2}$:
\begin{eqnarray}
\mathcal{M}^{D_s^-{D}^{*0}\to D_s^-{D}^{*0}}_{\textbf{Cont}}&=&-d_2m_{D_s}m_{D^*}\epsilon_{2\mu}\epsilon_4^{\dag \mu},\\
\mathcal{M}^{D_s^-{D}^{*0}\to D_s^{*-}{D}^{0}}_{\textbf{Cont}}&=&-2d_3\sqrt{m_{D_s}m_{D^*}m_{D_s^*}m_{D}}\epsilon_{2\mu}\epsilon_3^{\dag \mu},\\
\mathcal{M}^{D_s^{*-}{D}^{0}\to D_s^{*-}{D}^{0}}_{\textbf{Cont}}&=&-d_2{m_{D^*_s}m_{D}}\epsilon_{1\mu}\epsilon_3^{\dag \mu},\\
\mathcal{M}^{D_s^-{D}^{*0}\to D_s^{*-}{D}^{*0}}_{\textbf{Cont}}&=&-\sqrt{m_{D_s}m_{D_s^*}}m_{D^*}(d_5p_1^\mu-d_6p_3^\mu\nonumber\\
&&-d_7p_4^\mu+d_8 p_2^\mu)\epsilon_{\mu\nu\alpha\beta}\epsilon_2^\beta \epsilon_3^{\dag \nu}\epsilon_4^{\dag \alpha},\\
\mathcal{M}^{D_s^{*-}{D}^{0}\to D_s^{*-}{D}^{*0}}_{\textbf{Cont}}&=&-\sqrt{m_{D}m_{D^*}}m_{D_s^*}(-d_5p_2^\mu+d_6p_4^\mu\nonumber\\
&&+d_7p_3^\mu -d_8 p_1^\mu)\epsilon_{\mu\nu\alpha\beta}\epsilon_1^\beta \epsilon_3^{\dag\alpha}\epsilon_4^{\dag \nu},\\
\mathcal{M}^{D_s^{*-}{D}^{*0}\to D_s^{*-}{D}^{*0}}_{\textbf{Cont}}&=&(-2d_9\epsilon_{1\mu}\epsilon_{2\nu}\epsilon_3^{\dag\mu}\epsilon_4^{\dag\nu}-2d_{10}\epsilon_{1\mu}\epsilon_{2\nu}\epsilon_3^{\dag\nu}\epsilon_4^{\dag\mu}\nonumber\\
&&-2d_{11}\epsilon_{1\mu}\epsilon_{2}^\mu\epsilon_{3\nu}^{\dag}\epsilon_4^{\dag\nu})m_{D_s^*}m_{D^*}.
\end{eqnarray}
The amplitudes corresponding to the t-channel diagrams intermediated by $\eta$ and $\eta^\prime$ are shown as follows
\begin{eqnarray}
\mathcal{M}_{\eta^{(\prime)}}^{D_s^{*-}{D}^{0}\to D_s^{*-}{D}^{0}}&=&-\delta\frac{2g^2}{3F_\pi^2}\sqrt{m_{D_s}m_{D_s^*}m_Dm_{D^*}}q^\mu q^\nu \epsilon_{2\nu}\epsilon_{3\mu}^\dag\nonumber\\
&&\times\frac{1}{q^2-m_{\eta^{(\prime)}}^2+i\epsilon},\\
\mathcal{M}_{\eta^{(\prime)}}^{D_s^-{D}^{*0}\to D_s^{*-}{D}^{0}}&=&-\delta \frac{2g^2}{3f_\pi^2}\sqrt{m_{D_s}m_{D_s^*}m_Dm_{D^*}}q^\mu q^\nu \epsilon_{2\nu}\epsilon_{3\mu}^\dag\nonumber\\
&&\times\frac{1}{q^2-m_{\eta^{(\prime)}}^2+i\epsilon},\\
\mathcal{M}^{D_s^-{D}^{*0}\to D_s^{*-}{D}^{*0}}_{\eta^{(\prime)}}&=&-\delta i\frac{g^2\sqrt{m_{D_s}m_{D_s^*}}}{3F_\pi^2}q_\mu \epsilon_3^{\dag \mu} \epsilon^{\mu^\prime \nu \alpha \beta}(p_{2\mu^\prime}\nonumber\\
&&+p_{4\mu^\prime})q_\alpha \epsilon_{2\nu} \epsilon_{4\beta}^\dag \frac{1}{q^2-m_{\eta^{(\prime)}}^{2}+i\epsilon},\label{eq21}\\
\mathcal{M}^{D_s^{*-}{D}^{0}\to D_s^{*-}{D}^{*0}}_{\eta^{(\prime)}}&=&\delta i\frac{g^2\sqrt{m_{D}}m_{D_s^*}}{3F_\pi^2 \sqrt{m_{D^*}}}\epsilon^{\mu \nu \alpha \beta}(p_{1\mu}+p_{3\mu})\nonumber\\
&&\times \epsilon_{1\beta} \epsilon_{3\nu}^\dag\epsilon_{4\mu^\prime}^{\dag }q_\alpha q^{\mu^\prime}\frac{1}{q^2-m_{\eta^{(\prime)}}^{2}+i\epsilon},\label{eq22}\\
\mathcal{M}^{D_s^{*-}{D}^{*0}\to D_s^{*-}{D}^{*0}}_{\eta^{(\prime)}}&=&-\delta\frac{g^2}{6F_\pi^2}\epsilon^{\mu\nu\alpha\beta}\epsilon^{\mu^\prime\nu^\prime\alpha^\prime\beta^\prime}(p_{1\mu}+p_{3\mu})\nonumber\\
&&\times(p_{2\mu^\prime}+p_{4\mu^\prime})\epsilon_{1\beta}\epsilon_{2\nu^\prime}\epsilon_{3\nu}^\dag\epsilon_{4\beta^\prime}^\dag q_\alpha q_{\alpha^\prime}\nonumber\\
&&\times\frac{1}{q^2-m_{\eta^{(\prime)}}^{2}+i\epsilon}\label{eq26}.
\end{eqnarray}
In Eqs. (\ref{eq21}, \ref{eq22}, \ref{eq26}), $\delta=+1$ for the case of $\eta$ exchange, and $\delta=-1$ for $\eta^\prime$ exchange.

The partial-wave amplitude in the $lSJI$ basis for the transition $(lSJI)\to (l^\prime S^\prime) JI$ is
\begin{eqnarray}
&&\mathcal{M}^{JI}_{lS;l^\prime S^\prime}(s)\nonumber\\
&=&\frac{Y_{l0}(\hat{\textbf{z} })}{(2J+1)}\sum_{\lambda_1,\lambda_2, \lambda_3,\lambda_4, m^\prime}\langle s_3,\lambda_3;s_4,\lambda_4|S^\prime,\lambda_3+\lambda_4\rangle \nonumber\\
&&\times\langle s_1,\lambda_1;s_2,\lambda_2|S,\lambda_1+\lambda_2\rangle \langle l,0;s,\lambda_1 +\lambda_2|J,\lambda_1+\lambda_2\rangle\nonumber\\
&&\times\langle l^\prime,m^\prime;s^\prime,\lambda_3+\lambda_4|J,\lambda_1+\lambda_2\rangle \int d\hat{\textbf{p}}^\prime Y_{l^\prime m^\prime}(\hat{\textbf{p}}^\prime)^*\nonumber\\
&&\times M^I(p\hat{\textbf{z}},s_1,\lambda_1,s_2,\lambda_2;\textbf{p}^\prime,s_3,\lambda_3,s_4,\lambda_4).\label{eq27}
\end{eqnarray}
In this paper, we only consider S wave contribution, i.e., $l=l^\prime =0$.

The partial wave projection Eq. \eqref{eq27} for t-channel exchange amplitude would develop a left-hand cut by the following formulation \cite{Gulmez:2016scm, Du:2018gyn}
\begin{eqnarray}
A^{ij}(m_{ex}^2,s)&=&\int_{-1}^1d\cos\theta \frac{1}{q^2-m_{ex}^2+i\epsilon}\nonumber\\
&=&\frac{1}{2pp^\prime}\log \left(\frac{2pp^\prime +t_s^{ij}-m_{ex}^2+i\epsilon}{-2pp^\prime +t_s^{ij}-m_{ex}^2+i\epsilon}\right),\label{eq27p}
\end{eqnarray}
with 
\begin{eqnarray}
t_s^{ij}=\frac{m_1^2+m_2^2+m_3^2+m_4^2-s}{2}+\frac{(m_4^2-m_3^2)(m_1^2-m_2^2)}{2s},
\end{eqnarray}
where $p$ and $p^\prime$ are the momenta of each initial and final particles in the center-of-mass frame, respectively, $m_1$ and $m_2$ denote the masses of the initial particles, $m_3$ and $m_4$ the masses of the final ones, and $m_{ex}$ is the mass of the intermediate meson. Note that if the momentum $p$ or $p^\prime$ is zero, the value of $A^{ij}(m_{ex}^2,s)$ is defined as the limit of the result in Eq. (\ref{eq27p}), which is given by $\frac{2}{t_s^{ij}-m_{ex}^2}$.

Finally, the partial-wave amplitudes are given by
\begin{eqnarray}
V_{11}&=&d_2 m_{D_s}m_{D^*},\label{eq29}\\
V_{12}&=&\frac{\sqrt{m_Dm_{D_s}m_{D^*}m_{D_s^*}}}{9f_\pi^2}(18d_3f_\pi^2\nonumber\\
&&+g^2A^{12}(m_\eta^2,s)(m_\eta^2-p_1^2-p_2^2-t_s^{12})\nonumber\\
&&+g^2A^{12}(m_{\eta^\prime}^2,s)(-m_{\eta^\prime}^2+p_1^2+p_2^2+t_s^{12})),\\
V_{13}&=&\frac{\sqrt{2m_{D_s}m_{D_s^*}}m_{D^*}}{9f_\pi^2}(9i\bar{d}f_\pi^2\nonumber\\
&&+g^2A^{13}(m_\eta^2,s)(m_\eta^2-p_1^2-p_3^2-t_s^{13})\nonumber\\
&&+g^2A^{13}(m_{\eta^{\prime}}^2,s)(-m_{\eta^{\prime}}^2+p_1^2+p_3^2+t_s^{13})),\\
V_{22}&=&d_2 m_D m_{D_s^*},\\
V_{23}&=&\frac{\sqrt{2}m_Dm_{D_s^*}}{9f_\pi^2\sqrt{m_Dm_{D^*}}}(9i\bar{d}f_\pi^2m_{D^*}\nonumber\\
&&+g^2m_{D^*_s}A^{23}(m_\eta^2,s)(m_\eta^2-p_2^2-p_3^2-t_s^{23})\nonumber\\
&&+g^2m_{D^*_s}A^{23}(m_{\eta^{\prime}}^2,s)(-m_{\eta^{\prime}}^2+p_2^2+p_3^2+t_s^{23})),\\
V_{33}&=&\frac{m_{D^*}m_{D_s^{*}}}{9f_\pi^2}(18(-d_9+d_{10})f_\pi^2\nonumber\\
&&+g^2A^{33}(m_\eta^2,s)(m_\eta^2-2p_3^2-t_s^{33})\nonumber\\
&&+g^2A^{33}(m_{\eta^{\prime}}^2,s)(-m_{\eta^{\prime}}^2+2p_3^2+t_s^{33})) , \label{eq34}
\end{eqnarray}
where $p_i$ is the three momentum of the channel $i$. From these potentials we clearly see that the contributions of $\eta$ and $\eta^\prime$ exchange cancel each other, so the potentials corresponding to the diagrams of t-channel are very small.

\section{The Bethe-Salpeter Equation in The On-shell factorized Form}
With the interaction potentials obtained above, we use the coupled channel Bethe-Salpeter equation in the on-shell factorized form to obtain the $T$-matrix \cite{Oller:1997ti}, i.e.,
\begin{equation}
T^J=(I-V^JG)^{-1}V^J,
\end{equation}
where $V^J$ is the partial-wave amplitude $\mathcal{M}^{JI}_{lS;l^\prime S^\prime}(s)$ obtained above, and $G$ is the two-meson loop function, given by
\begin{equation}
G = i\int \frac{d^4q}{(2\pi)^4}\frac{1}{q^2-m_1^2+i\epsilon}\frac{1}{(P-q)^2-m_2^2+i\epsilon}
\end{equation}
with $P^2=s$, $m_1$ and $m_2$ the particle masses of the corresponding channel. This loop integral is logarithmically divergent, and can be calculated with three-momentum cutoff regularization. The  analytic expression of the cutoff regularization for the loop function has been obtained in the Ref. \cite{Oller:1998hw}, i.e., 
\begin{eqnarray}
G&=&\frac{1}{32\pi^2}\left\{\frac{\nu}{s}\left[\log\frac{s-\Delta+\nu\sqrt{1+\frac{m_1^2}{q_{max}^2}}}{-s+\Delta+\nu\sqrt{1+\frac{m_1^2}{q_{max}^2}}}\right.\right.\nonumber\\
&&\left.+\log\frac{s+\Delta+\nu\sqrt{1+\frac{m_1^2}{q_{max}^2}}}{-s-\Delta+\nu\sqrt{1+\frac{m_1^2}{q_{max}^2}}}\right]-\frac{\Delta}{s}\log \frac{m_1^2}{m_2^2}\nonumber\\
&&+\frac{2\Delta}{s}\log\frac{1+\sqrt{1+\frac{m_1^2}{q_{max}^2}}}{1+\sqrt{1+\frac{m_2^2}{q_{max}^2}}}+\log\frac{m_1^2m_2^2}{q_{max}^4}\nonumber\\
&&\left.-2\log\left[\left(1+\sqrt{1+\frac{m_1^2}{q_{max}^2}}\right)\left(1+\sqrt{1+\frac{m_2^2}{q_{max}^2}}\right)\right]\right\},\label{eq37}
\end{eqnarray}
where $q_{max}$ stands for the cutoff, $\Delta=m_2^2-m_1^2$, and $\nu=\sqrt{[s-(m_1+m_2)^2][s-(m_1-m_2)^2]}$. The cutoff regularization enables us to evaluate the loop function with $q_{max}$ around 1 GeV, which is a natural value \cite{Oller:2000fj}.

Note that Eq. (\ref{eq37}) justifies in the physical sheet, also called the first Riemann sheet. To look for the poles, we need to extrapolate the loop function of Eq. \eqref{eq37} to the second Riemann sheet by an continuation via 
\begin{equation}
G^{II}=G^I+i\frac{\nu(s)}{8\pi s},
\end{equation}
where $\nu(s)=\sqrt{(s-(m_1+m_2)^2)(s-(m_1-m_2)^2)}$.

\section{Numerical Results and Discussion}

In order to make the numerical calculation, we need to evaluate the coupling constants appearing in the effective potentials, i.e., Eqs. (\ref{eq29}-\ref{eq34}). For $a_1$, $b_1$, $c_2$ and $c_3$, we compare the Lagrangians in Eqs. (\ref{eq1}) with the ones in Ref. \cite{Ding:2008gr, Casalbuoni:1996pg, Sun:2011uh}, and we obtain
\begin{eqnarray}
&&a_1=-\frac{\beta g_V}{m_{D}g},b_1=2g,c_2=-\frac{g}{m_{D^*}},c_3=\frac{\beta g_V}{m_{D^*}g}.
\end{eqnarray}
For $d_9$, $d_{10}$, $d_{11}$, $e_1$, $e_2$, $h$, $k_1$ and $k_2$, we assume that these constants are saturated by the resonances, such as the pseudoscalar, vector and axial-vector mesons, so that the following relations are obtained
\begin{eqnarray}
&&d_9=-\frac{1}{2}\left(\frac{2e_2^2}{m_{a_0}^2}-\frac{4e_1^2}{m_{S_1^2}}-\frac{4h^2}{4m_{D^{*2}}-m_{a_1}^2}\right),\nonumber\\
&&d_{10}=-\frac{1}{2}\left(-\frac{2h^2}{m_{a_1}^2}+\frac{2h^2}{3m_{f_1^2}}+\frac{4e_2^2}{4m_{D^{*2}}-m_{a_0}^2}\right),\nonumber\\
&&d_{11}=-\frac{1}{2}\left(\frac{2h^2}{m_{a_1}^2}-\frac{2h^2}{3m_{f_1^2}}+\frac{4h^2}{4m_{D^{*2}}-m_{a_1}^2}\right),\nonumber\\
&&e_2=-\frac{g_{DDS}}{2\bar{m}_{D}},e_1=\frac{e_2}{\sqrt{3}},h^2=e_2^2.
\end{eqnarray}
Furthermore, we estimate that $(d_5m_{D_s}-d_6m_{D_s^*}-d_7m_{D^*}+d_8m_{D})^2\sim \bar{d}^2$ and $(d_5m_{D}-d_6m_{D^*}-d_7m_{D_s^*}+d_8m_{D_s^*})^2\sim \bar{d}^2$ where $\bar{d}^2\sim 7.5\times 10^{-12}$ MeV$^{-4}$.
The values of $g$, $g_{DDS}$, $g_V$, $\beta$, $\lambda$ and $g_1$ are taken from Refs. \cite{Sun:2011uh, Du:2016tgp}, respectively, i.e.,
\begin{eqnarray}
&&g=0.59,\ g_V=5.8,\ \beta=0.9, \nonumber\\
&&\lambda=0.56\ \mathrm{GeV}^{-1},\ g_{DDS}=5058\ \mathrm{MeV}.
\end{eqnarray}

\begin{figure}
\centering
\includegraphics[width=0.85\linewidth]{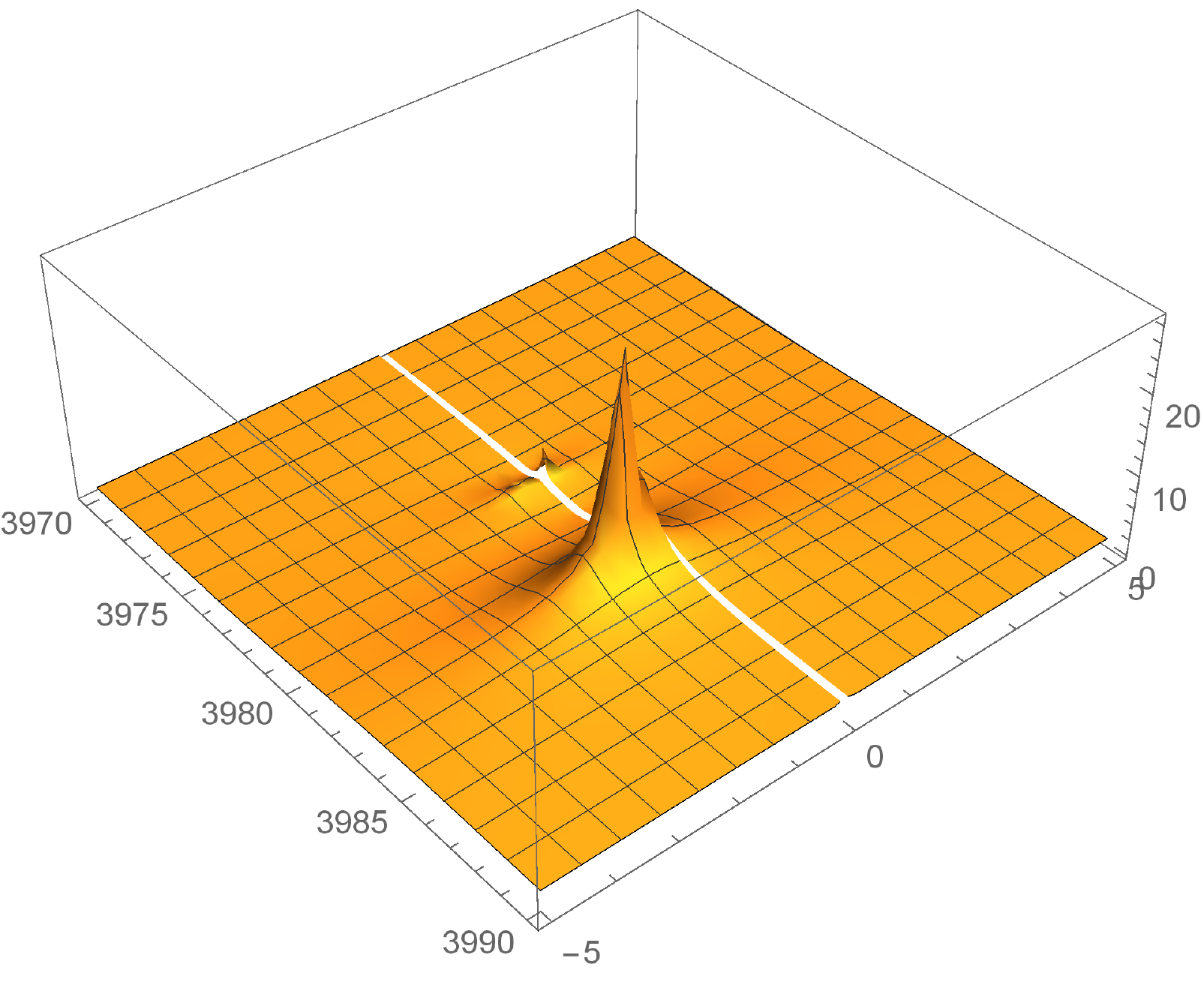}
\caption{Three dimension plot of the pole positionstructure on the complex plane.\label{fig1}}
\end{figure}

With the effective potentials under partial wave projection, we solve the Bethe-Salpeter equation taking into account the $D_s^+\bar{D}^*$, $D_s^*\bar{D}$ and $D_s^*\bar{D}^*$ coupled channels. We find a pole on the third Riemann sheet whose location is around $s=3985 \pm 0.5$ MeV. Note that the cut off $q_{max}$ should be chosen of the order of magnitude of $1$ GeV as usual. If this cut off goes from $900$ MeV to 1300 MeV, the pole position moves from $3981.57-i0.48$ MeV to $3984.90-i0.27$ MeV. In Table \ref{tab1}, we list the pole positions corresponding to different values of $q_{max}$, where once can see that our results are not much depended on the cutoff. 
\begin{table}
\caption{Poles for varying the cut off with the unit of MeV.}\label{tab1}
\begin{tabular}{|c|c|c|c|ccccccccc}\toprule[1pt]
$q_{max}$&900&950&1000\\
Pole &3981.57-i0.48&3981.95-i0.51&3982.34-i0.53\\\midrule[0.5pt]
$q_{max}$&1050&1100&1150\\
Pole &3982.74-i0.53&3983.16-i0.514805&3983.59-i0.48\\\midrule[0.5pt]
$q_{max}$&1200&1250&1300\\
Pole &3984.02-i0.43&3984.46-i0.36&3984.90-i0.27\\\bottomrule[1pt]
\end{tabular}
\end{table}
In Fig. \ref{fig1}, we plot the 3 dimension structure of the pole position on the complex plane with the typical value of $q_{max}=1000$ MeV.
Since the $Z_{cs}^-(3985)$ has a mass of $3982.5^{+1.8}_{-2.6}\pm 2.1$ MeV and a width of $12.8^{+5.3}_{-4.4}\pm 3.0$ MeV, which is consistent with our results. 
Note that, this pole is located below the threshold of the $D_s^*\bar{D}^*$ channel. But, we have checked that the potential of the $D_s^*\bar{D}^*$ channel, see Eq. \eqref{eq34}, is not strong enough to reproduce a bound state when only a single channel interaction is taken into account. Therefore, the coupled channel effects of the $D_s^+\bar{D}^*$ and $D_s^*\bar{D}$ channels play much important role to create the pole.
Thus, we can conclude that this newly observed state is a $D_s^{(*)-}D^{(s*)0}$ molecular state, which is a loose bound state of $D_s^*\bar{D}^*$ with significant components of $D_s^+\bar{D}^*$ and $D_s^*\bar{D}$, analogously to the one of $a_0(980)$ as commented in Ref. \cite{Ahmed:2020kmp}.

In addition, we need to mention that the above results are obtained choosing the sign of $\bar{d}$ as negative. However, this sign can not be fixed in our theoretical frame. If we chose the positive sign, the results dose not change much. The reason is that the contribution of the contact term of $D_s^-D^{*0}\to D_s^{*-}D^{*0}$ ($D_s^{*-}D^{0}\to D_s^{*-}D^{*0}$) dose not change much the rest part of the potential $V_{13}$ ($V_{23}$), since the modulus of the contact term is about 3-10 times smaller than the rest part of the corresponding potential in the energy region of 3980-3988 MeV. 
The location of the pole moves from $3981.55-i0.68$ MeV to $3984.89-i0.72$ MeV, as the cut off changes from 900 MeV to 1300 MeV. In such a case, our conclusion is unchanged that the $Z_{cs}^-(3985)$ can be explained as a $D_s^{(*)-}D^{(*)0}$ molecular state.

\section{Summary}
In this work, we construct the Lagrangians of charmed mesons and pseudoscalar/vector mesons as well as the contact terms describing the interactions of four charmed mesons, considering the chiral symmetry and the hidden local symmetry. 

With the Lagrangians constructed, we calculation the effective potentials under the partial wave projection, and then solve the Bethe-Salpeter equation with the on-shell factorization. On the third Reimann sheet of the complex plane, we get a pole which locates at $3982.34-i0.53$ MeV choosing the cut off as 1000 MeV. If changing the cut off in the region of 900-1300 MeV, the real part of the position move less than 4 MeV, and the imaginary part moves within 0.3 MeV. Our results can explain very well the $Z_{cs}^-(3985)$ reported by recent BESIII experiment, which implies that the $Z_{cs}^-(3985)$ can be explained as a $D_s^{(*)-}D^{(*)0}$ molecular state.

\section*{Acknowledgments}
This project is suported by the National Natural Science Founadtion of China (NSFC) under Grants No. 11705069 and 11965016.

\bibliographystyle{unsrt}

\end{document}